\documentclass[twocolumn,superscriptaddress,nofootinbib,floatfix,aps,prb,citeautoscript,longbibliography,10pt]{revtex4-2}
\usepackage[utf8]{inputenc}
\usepackage[T1]{fontenc}
\usepackage{lineno}

\usepackage{graphicx}
\usepackage{color}
\usepackage{array}
\usepackage{dcolumn}
\usepackage{bm}
\usepackage{multirow}
\usepackage[version=4]{mhchem}
\usepackage{booktabs}
\usepackage{xr-hyper}
\usepackage{hyperref}
\usepackage{cleveref}
\usepackage{enumitem}
\usepackage{framed}
\usepackage{siunitx}
\usepackage{subcaption}
\definecolor{shadecolor}{RGB}{225,225,225}

\usepackage[usenames,dvipsnames]{xcolor} 

\begin{document}
\newcommand{\ETH}{Department of Materials, ETH Zürich, Zürich, CH-8093, Switzerland}
\newcommand{\UV}{Faculty of Physics, University of Vienna, Vienna, 1090, Austria}
\newcommand{\UW}{Department of Chemistry, University of Warwick, Coventry, CV4 7AL, United Kingdom}
\newcommand{\RUB}{Interdisciplinary Centre for Advanced Materials Simulation, Ruhr University Bochum, D-44799 Bochum, Germany}
\newcommand{\IRB}{Ru{\dj}er Bo\v{s}kovi\'{c} Institute, HR-10000 Zagreb, Croatia}
\newcommand{\UICE}{The Faculty of Industrial Engineering, Mechanical Engineering, and Computer Science, University of Iceland, Reykjavík, Iceland}
\newcommand{\UoM}{Department of Civil Engineering and Management, Faculty of Science and Engineering, The University of Manchester, Manchester, M13 9PL, United Kingdom}
\newcommand{\UBE}{Vin\v{c}a Institute of nuclear sciences-national institute of the Republic of Serbia, University of Belgrade, Serbia}
\newcommand{\INESCTEC}{Institute for Systems and Computer Engineering, Technology and Science, 4200-465 Porto, Portugal}
\newcommand{\UP}{School of Economics and Management, University of Porto, 4200-464 Porto, Portugal}
\newcommand{\PP}{CEOS.PP, ISCAP, Polytechnic of Porto, 4465-004 S. Mamede de Infesta, Portugal}
\newcommand{\CNRNano}{CNR-NANO S3, 41125 Modena, Italy}
\newcommand{\UNIMOREFIM}{Dipartimento di Scienze Fisiche, Informatiche e Matematiche, Universit\`a degli Studi di Modena e Reggio Emilia, 41125 Modena, Italy}
\newcommand{\EPFL}{Laboratory of Computational Science and Modeling, Institut des Mat\'eriaux, \'Ecole Polytechnique F\'ed\'erale de Lausanne, 1015 Lausanne, Switzerland}
\newcommand{\TUD}{Department of Materials Science and Engineering, Delft University of Technology, Delft, Netherlands}
\newcommand{\cass}{Climate Safety and Security Centre, TU Delft The Hague Campus, Delft University of Technology, 2594 AC, The Hague, The Netherlands}
\newcommand{\VU}{Vilnius University, Life Sciences Center, Institute of Biotechnology, Saul\.{e}tekio al. 7, LT-10257 Vilnius, Lithuania}
\newcommand{\UTU}{Department of Mechanical and Materials Engineering,  University of Turku, Turku 20014, Finland}
\newcommand{\Syngenta}{Syngenta Crop Protection AG, Schaffhauserstrasse, Stein, 4332, AG, Switzerland}
\newcommand{\NOMATEN}{NOMATEN Centre of Excellence, National Center for Nuclear Research, ul. A. Sołtana 7, 05-400 Swierk/Otwock, Poland.}
\newcommand{\MUG}{NanoLab, Division of Medicinal Chemistry, Otto Loewi Research Center, Medical University of Graz, Neue Stiftingtalstraße 6, 8010, Graz, Austria}
\newcommand{\BTM}{BioTechMed-Graz, Mozartgasse 12, 8010 Graz, Austria}
\newcommand{\SU}{Sabanci University, Orta Mah, FENS 1023, Tuzla, Istanbul, Turkey}

\author{Lukas H\"{o}rmann}
\affiliation{\UV}
\affiliation{\UW}

\author{Hemanadhan Myneni}
\affiliation{\UICE}

\author{Rwayda Kh. S. Al-Hamd}
\affiliation{\UoM}

\author{Katarina Batalovi\'{c}}
\affiliation{\UBE}

\author{Silvia Bonfanti}
\affiliation{\NOMATEN}

\author{Federico Grasselli}
\affiliation{\UNIMOREFIM}
\affiliation{\CNRNano}

\author{Saulius Gra\v{z}ulis}
\affiliation{\VU}

\author{Bahattin Ko\c{c}}
\affiliation{\SU}

\author{Konstantinos Konstantinou}
\affiliation{\UTU}

\author{Ivor Lon\v{c}ari\'{c}}
\affiliation{\IRB}

\author{Nataliya Lopanitsyna}
\affiliation{\Syngenta}

\author{Jos\'{e} Manuel Oliveira}
\affiliation{\INESCTEC}
\affiliation{\UP}

\author{Paolo Pegolo}
\affiliation{\EPFL}

\author{Patr\'{i}cia Ramos}
\affiliation{\INESCTEC}
\affiliation{\PP}

\author{Kevin Rossi}
\affiliation{\TUD}
\affiliation{\cass}

\author{Sebastian P. Schwaminger}
\affiliation{\MUG}
\affiliation{\BTM}

\author{Edith Simmen}
\affiliation{\ETH}

\author{Milica Todorovi\'c}
\affiliation{\UTU}

\author{Markus Stricker}
\thanks{Corresponding author: markus.stricker@rub.de}
\affiliation{\RUB}

\author{Jonathan Schmidt}
\affiliation{\ETH}
\affiliation{\EPFL}

\date{\today}

\title{Journal Research Data Policies in Materials Science}

\begin{abstract}
Open and reproducible research in materials science relies on the availability of data, code, and common metadata standards. Journal research data policies (RDPs) remain a primary mechanism by which publication norms are defined and enforced. We survey RDPs for 171 materials science journals spanning 17 publishers, using an expanded coding framework that captures both data-and-code sharing behavior as well as refereeing standards. We find clear signs of progress in comparison to earlier research on RDPs: nearly all journals provide an RDP, and most mention data availability statements. However, enforceable requirements remain uncommon, public deposition of underlying data is rarely mandatory, and FAIR publication is typically encouraged rather than required. Expectations for research software are substantially less developed than those for data, with limited attention to versioning and persistent identifiers, dependency disclosure, reproducible execution environments, or software quality practices. Aggregating the findings on policy features into an open research data score reveals pronounced heterogeneity across journals. Neither impact factor nor access model reliably predicts policy strength. Double-coding further shows that more complex policies and stricter policies can be more challenging to interpret consistently, and we highlight challenges in consistent RDP encoding across studies. Lastly, we conclude with recommended best practice directions for the future.
\end{abstract}

\maketitle

\section{Introduction}

The field of materials science has witnessed a transformative acceleration of research driven by the integration of data-driven approaches and machine learning~\cite{Tanaka2018,Sandfeld2018,Scheffler2022,Bauer2024}. 
As researchers strive to develop innovative materials with tailored properties for applications ranging from energy storage and conversion to electronics by relying on data-heavy workflows, the need for data-availability has never been more critical. Achieving verifiable results requires systematic management and sharing of the underlying research artifacts, including all datasets and the software and workflows used to generate and analyze them, beyond what can be reported in the manuscript alone~\cite{Kononova2021,Schilling-Wilhelmi2025,Wilkinson2016}.
An environment of data sharing, code sharing and competitive benchmarks ideally lead to a state of \textit{frictionless reproducibility} which has been a key driver of progress in machine learning~\cite{singularity} and its application to materials science.

A commitment to open research practices fosters collaboration, accelerates discovery, enables validation of results, and facilitates the reuse of tools across disciplines~\cite{Scheffler2022,Ramachandran2021}. For data-driven methods in particular, the availability of high-quality, machine-readable datasets forms the essential foundation for building robust models~\cite{Shumailov2024}.   

Data sharing practices vary substantially across scientific disciplines. In crystallography and structural biology, data sharing is well established: the Protein Data Bank (PDB) founded in the 1970s~\cite{ProteinDataBank1971} evolved into the global archive for macromolecular structures~\cite{Bernstein1977,Westbrook2022}. By 2008, deposition of not only structure coordinates but also structure factors and Nuclear Magnetic Resonance restraints became mandatory~\cite{Terwilliger2014}, enabling validation and, in some cases, improvement of the original findings. This policy, coordinated by the Worldwide PDB consortium~\cite{wwPDB2024} and supported by major publishers and funding agencies, substantially improved the availability of structural data and enabled machine learning success stories, such as in protein folding prediction~\cite{alphafold0, alphafold1, alphafolddb, prealphafold}. More recently, advances in computing and storage have enabled repositories for raw diffraction and microscopy data~\cite{natureEditorial2016}.

In the materials science community, data sharing is likewise recognized as essential to advancing the field, as highlighted by the adoption of international frameworks such as the FAIR (Findable, Accessible, Interoperable, Reusable) principles for data~\cite{Wilkinson2016} and code~\cite{Barker2022} in materials science~\cite{Ghiringhelli2023}. Although no single centralized repository exists for crystal structure data, maintainers of major computational databases~\cite{aflow, materialsproject, oqmd, nomad, alexandria, materials-cloud} have integrated the OPTIMADE API standard~\cite{Evans2024}, effectively creating a common data space for basic, simulated crystal structure data. Although crystal structures are only a small subspace of the diverse materials science research data, these successful data-sharing initiatives demonstrate that best practices can be adapted to address gaps in materials-data sharing standards.

Various ongoing projects and communities work on the development of data and metadata standards, such as NOMAD~\cite{nomad}, the Battery Data Genome~\cite{Ward2022}, the Materials Genome Initiative~\cite{White2012}, and the Materials Research Data Alliance~\cite{marda2025}, and strategies have been proposed to encourage authors to share data at all levels. This includes initiatives by funding agencies~\cite{ERC2022,Forschungsgemeinschaft2025,SNSF2025,CSF,Rannis_Iceland,tdcc_nes}, universities~\cite{UniRuhrBochum2018,UniVilnius2022}, research communities~\cite{Mansour2023}, and individual research groups.

Despite global community efforts, issues such as the lack of standardization of data formats, reluctance to share proprietary or sensitive data, insecurities about sharing data and code, and insufficient infrastructure for storage and access continue to hinder progress~\cite{Himanen2019, Gomes2022, Ward2022}. Furthermore, the complexity, heterogeneity, and sheer volume of materials science data --- much of it generated without established curation frameworks --- pose persistent obstacles to effective data management~\cite{Hartl2021}. In combination with the rapid expansion of machine-learning–driven research, these limitations raise a second, more systemic risk: low-quality, weakly documented, or insufficiently validated datasets and workflows can be reused at scale, silently propagating errors through follow-on studies and into widely used databases, generating noise and erroneous conclusions. Strengthening data stewardship and reproducibility practices is therefore not only a matter of more efficient data management, but also a prerequisite for maintaining credibility as machine learning (ML) becomes increasingly embedded in materials science.

Research across STEM disciplines (science, technology, engineering, mathematics) highlights that journals are the arbiters of community practice and can, even more than funding bodies, incentivize authors to follow community-specific best practices~\cite{Kim2016a} that move the field forward~\cite{stodden2013toward, vasilevsky2017reproducible, mccain1995mandating, barbui2016sharing}.
However, such journal-specific policies have often been found inadequate in their standards, and in some cases, entirely absent~\cite{Resnik2019, rousi2020journal, Crosas2018, vasilevsky2017reproducible, Castro2017, Naughton2016, blahous2016data, Herndon2016, Sturges2015, ZenkMltgen2014, Moles2015, gherghina2013, Weber2010,Piwowar2008}. For example, Rousi et al. (2020)~\cite{rousi2020journal} reported that \SI{35}{\percent} of the highly cited physics journals they examined had no research data policy (RDP), while Resnik et al. (2019)~\cite{Resnik2019} found that \SI{43.6}{\percent} of the examined journals did not address data sharing in their policies at all. 

There are systemic challenges to adopting more comprehensive and stringent RDPs. First, such policies inevitably impose additional workload on researchers, reviewers, and editors~\cite{GrantHrynaszkiewicz2018}, which can conflict with the incentives of profit-driven journals. Stricter standards may slow publication processes and reduce the number of accepted manuscripts, potentially lowering revenue, particularly in open-access models where income is tied to publication volume. Second, in many materials science domains, companies are significant contributors to research but are often unable to share data~\cite{Suhr2020}. Enforcing strict data publication standards could reduce their scientific contributions to public research. On the other hand, access to results and code is necessary for referees and editors to make informed judgments about reproducibility. Currently, it is largely up to individual editors and reviewers to insist on good practices prior to acceptance, but their mandate is limited without explicit journal policies related to data and code sharing.
On a positive note, there are already activities aiming at automating the process of good code sharing practices, such as FACILE-RS~\cite{Houillon2025}.

Nevertheless, longitudinal studies in other fields indicate progress in the adaptation of RDPs: in mathematical and multidisciplinary science journals, the fraction without data policies was \SI{66}{\percent} in 2011 and \SI{62}{\percent} in 2012~\cite{stodden2013toward}, while in the social sciences it decreased from \SI{90}{\percent} in 2003 to \SI{61}{\percent} in 2015. A comprehensive overview of these and related studies is provided in Ref.~\cite{rousi2020journal}.

In this article, we extend existing coding frameworks for journal research data policies~\cite{Resnik2019,rousi2020journal} to quantitatively assess the current policy landscape in materials science. Here, ``coding'' follows the social science meaning of systematically categorizing qualitative and quantitative information to enable structured analysis. Our adapted framework incorporates materials science–specific considerations, explicitly evaluates research software beyond code sharing (including expectations for code standards and computational workflow reproducibility), and captures enforcement mechanisms such as guidance for editors and referees (full framework in Sec.~\ref{sec:methods}). Using the Dimensions~\cite{dimensions} API and selected research field definitions, we compiled a dataset of 171 journals from 17 publishers that is broadly representative of materials science publication venues. For each publisher, we sampled --- where available --- the three highest impact journals (by journal impact factor), the three most prolific journals (by publication volume), and three randomly selected journals to test whether journal characteristics such as impact or popularity relate to policy strictness. We additionally included a small number of data-focused journals (e.g., \textit{CODATA Data Science Journal}) to benchmark best-practice-oriented policies. Lastly, we examine the challenges of consistently coding RDPs, both within a single study and when comparing results across different studies.

\section{Results}

\begin{figure*}
    \centering
    \includegraphics[width=\linewidth]{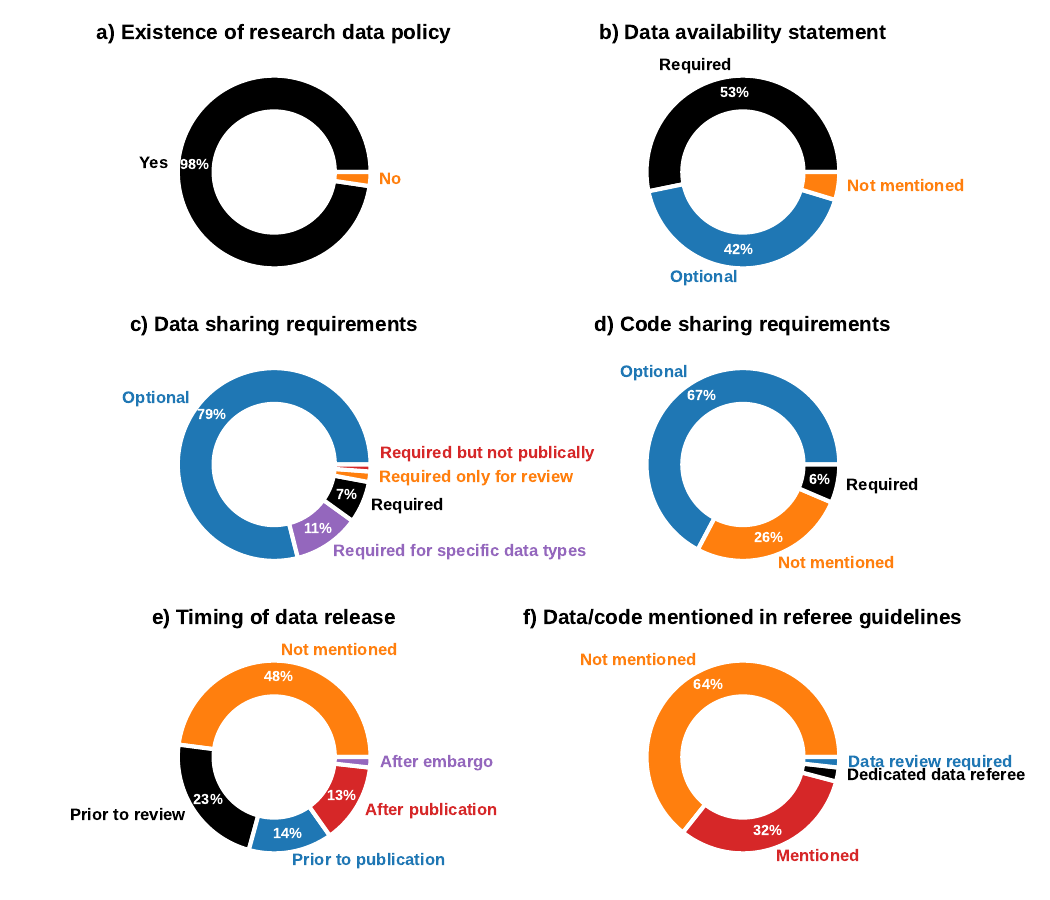}
    \caption{Overview of requirements for sharing data and code.}
    \label{fig:sharing_requirements}
\end{figure*}

An overview of the results concerning the fundamentals of the RDPs is given in Figure \ref{fig:sharing_requirements}. These results include a) the existence of an RDP in the first place, b) the rules concerning data availability statements (DAS), c) specifics on data sharing requirements, d) specifics on code sharing requirements, e) timing of data release, and f) whether data and/or code are mentioned in the journals' referee guidelines.

\noindent
\textbf{Data policies.}
To start, RDPs show a positive trend: compared to earlier studies~\cite{Resnik2019, rousi2020journal, stodden2013toward}, \SI{98}{\percent} of the journals we surveyed have an RDP (Figure~\ref{fig:sharing_requirements} a), and \SI{95}{\percent} mention a DAS (Figure~\ref{fig:sharing_requirements} b). However, nearly half of all journals still do not require a DAS. Although it is known that DAS alone are insufficient for establishing a strong data publishing culture~\cite{Gabelica2022, Hamilton2022}, we still consider it one of the foundations of an RDP, provided that the truthfulness of the DAS is verified during the review process before publication.
 
Most RDPs mention data sharing requirements, but typically as `optional' (Figure~\ref{fig:sharing_requirements} c): only \SI{7}{\percent} of journals require public sharing of all research data in all cases. Requirements for specific data types (e.g., crystal structures, genomics) are more common for \SI{11}{\percent} of the journals. An additional \SI{1.8}{\percent} require sharing of all data during peer review, and \SI{1.2}{\percent} require sharing upon reasonable request.

Since previous research has found that a DAS alone is not sufficient to foster good data-sharing behavior~\cite{Gabelica2022, Hamilton2022}, we also survey the timing of data release (cf.~Figure~\ref{fig:sharing_requirements} e) and the requirements to make data publicly available. Rousi et al.~\cite{rousi2020journal} found that 15 out of 40 surveyed physics journals required data to be released upon publication. In our materials-science sample, combining the categories ``required data available prior to the review process'' and ``required data available after official publication'', we find that \SI{36.3}{\percent} of journals include an explicit requirement for data availability by review and/or by publication. This is broadly comparable to, but slightly lower than, the corresponding fraction reported in Ref.~\cite{rousi2020journal}.

So far, we have only discussed whether and when research data are shared. Equally important is the location of the shared data, as this affects both its findability, long-term preservation, and adherence to metadata standards. Specifically, two-thirds of journals encourage sharing data on FAIR repositories, \SI{11.9}{\percent} require the sharing of certain data types on FAIR repositories, and only \SI{1.2}{\percent} require all data to be shared on FAIR repositories. Conversely, \SI{77.4}{\percent} of journals do not explicitly mention the licensing of data in their open data policies.
Following the FAIR definition~\cite{Wilkinson2016}, `Reusable' is defined as \textit{``(meta)data are released with a clear and accessible data usage license''}.
In any case, the mere existence of a license does not imply a permissible license and reusability per se. Hence, journal policies aimed at data sharing should support permissible licenses.

Strict policies provide little benefit without enforcement, which usually requires the expertise of referees. However, few journals consider the role of data in the peer review process, and \SI{61}{\percent} of journals do not mention the RDP in the referee guidelines (Figure~\ref{fig:sharing_requirements} f). Fewer than one-third make any reference to it, and only a very small number explicitly require data and/or code sharing in their referee guidelines or indicate that a dedicated data referee is appointed.

\noindent
\textbf{Code policies.}
In stark contrast to data policies, software code policies are weaker across all surveyed journals. Code sharing is encouraged but optional for \SI{70}{\percent} of journals and required in only \SI{1.8}{\percent}. Even among journals that encourage code sharing, more than \SI{80}{\percent} do not require authors to specify software dependencies or execution environments. Over \SI{80}{\percent} of policies omit guidance on persistent identifiers for code. Mentions of coding standards (e.g., style or safe constructs) are uncommon -- around \SI{80}{\percent} of policies do not address them -- and fewer than \SI{10}{\percent} require any minimal standards and references to testing and linting are rarer still.

\subsection*{Open Data Score}
\begin{figure}
    \centering
    \includegraphics[width=1\linewidth]{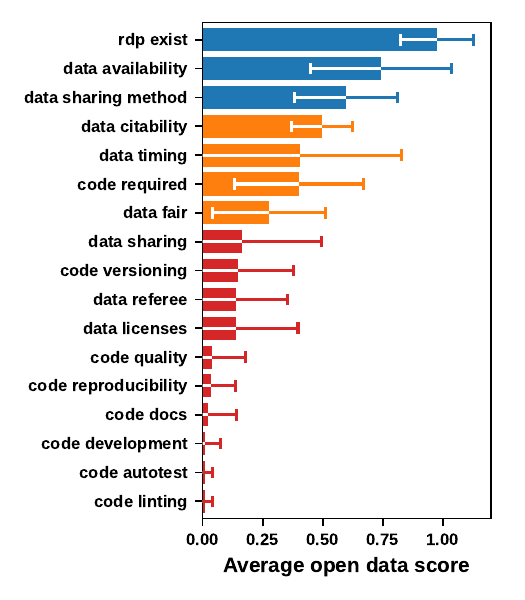}
    \caption{Average open data score by question; Error bars indicate standard deviation of open data scores. The colors indicate questions with ODS below 0.2 (red), below 0.5 (orange) and above 0.5 (blue).}
    \label{fig:score_vs_impact_average}
\end{figure}

To summarize policy strictness, we report an \emph{open data score} (ODS; \(s\in[0,1]\), higher = stricter; full definition in Section~\ref{sec:ods}). We evaluate the average ODS across journals, while the variability across journals for each question is captured by a per-question standard deviation. These results are presented in Figure~\ref{fig:score_vs_impact_average}. 

Journals achieve relatively high ODSs for the existence of an RDP, the requirements for a \textit{data availability} statement, and policies on \textit{data sharing methods}, with many journals recommending FAIR data repositories and \textit{data citability}. Conversely, concrete requirements for sharing data achieve lower ODSs. For instance, \textit{data sharing} requirements achieve an ODS of below $0.2$. Overall, we find large standard deviations. For instance, the large standard deviations for \textit{data availability} and \textit{data timing} indicate substantial differences in whether the respective DASs are required and when data must be released.

\begin{figure*}
    \centering
    \begin{subfigure}[t]{0.48\textwidth}
        \centering
        \includegraphics[width=\linewidth]{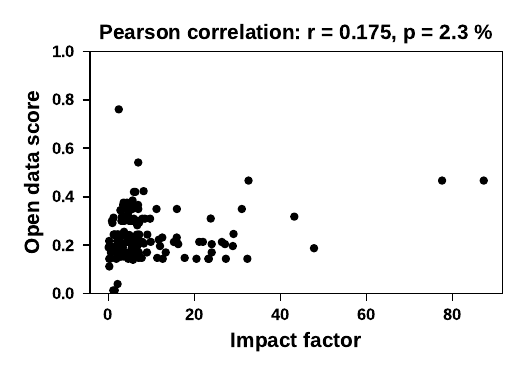}
        \caption{ODS versus impact factor of different journals.}
        \label{fig:score_vs_impact}
    \end{subfigure}
    \begin{subfigure}[t]{0.48\textwidth}
        \centering
        \includegraphics[width=\linewidth]{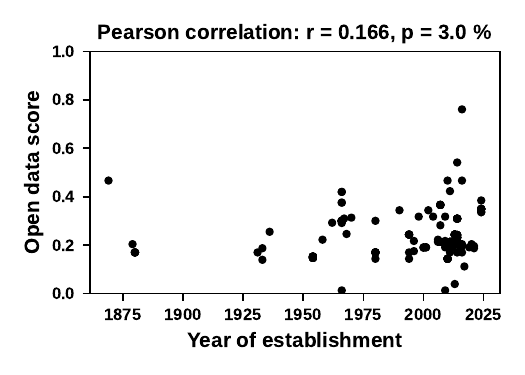}
        \caption{ODS versus year of the journals' establishment.}
        \label{fig:policy_vs_year_of_establishment}
    \end{subfigure}
    \begin{subfigure}[t]{0.48\textwidth}
        \centering
        \includegraphics[width=\linewidth]{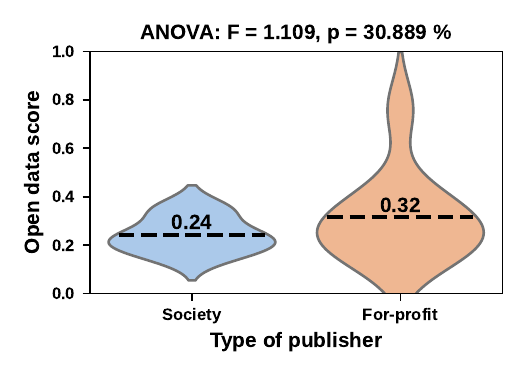}
    \caption{ODS versus type of publisher; the average ODS of each publisher is considered as one sample.}
        \label{fig:score_vs_society}   
    \end{subfigure}
    \begin{subfigure}[t]{0.48\textwidth}
        \centering
        \includegraphics[width=\linewidth]{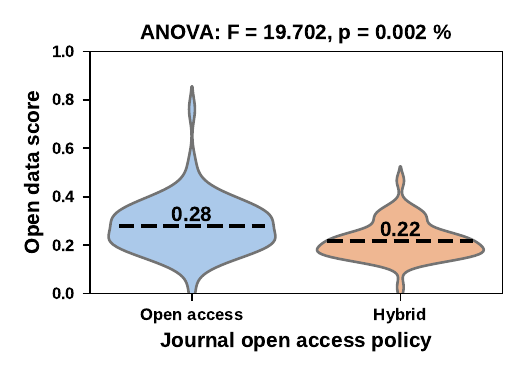}
    \caption{ODS versus open access policy of journals.}
    \label{fig:score_vs_open_access}
    \end{subfigure}
    \caption{Open data score (ODS) versus journal characteristics.}
    \label{fig:ods_vs_journal_characteristics}
\end{figure*}

We tested several hypotheses about factors associated with stricter RDPs, focusing on the relationships between the ODS and (i) journal impact factor, (ii) society vs.\ non-society publishers, (iii) open- vs.\ closed-access journals, and (iv) the year of the journal's establishment.

Figure~\ref{fig:score_vs_impact} illustrates the ODS versus journal impact factor. Although some high-impact journals have higher ODS, there is no clear monotonic trend: journals across the impact spectrum span the full ODS range, suggesting that prestige (impact factor) does not reliably correlate with policy strictness. The Pearson correlation is $r=0.175$ ($n=169$, $p=2.3\,\%$), indicating a weak but statistically still significant association.
To assess sensitivity to extreme values, we repeated the analysis after excluding the two highest-impact journals (pre-specified rule: impact factor $> 75$). The correlation disappeared ($r=-0.007$, $p=93.9\,\%$), indicating that the apparent association is driven by these outliers rather than a general trend.

Figure~\ref{fig:policy_vs_year_of_establishment} depicts the  ODS as a function of the year of establishment. The Pearson correlation is $r=0.166$ ($n=171$, $p=3.0\,\%$), indicating a weak but statistically significant association.

Figure~\ref{fig:score_vs_society} shows the relationship between the publisher type (society or for-profit) and the ODS. While we find that the average ODS of for-profit publishers is higher than that of society publishers, the ANOVA F-test as well as a t-test find no statistically significant correlations.

Figure~\ref{fig:score_vs_open_access} shows the relationship between the open access policy of a journal (full open access versus hybrid) and the ODS. We find that the average ODS of open-access journals is higher than that of journals with hybrid policies. To gauge significance, we perform an ANOVA F-test as well as a t-test. We find an F-statistic of \SI{19.702}. The p-score is \SI{0.002}{\percent}, indicating that this result is statistically significant.

\subsection*{Consistency of encoding process}
As detailed in Section~\ref{sec:encoding_process}, two researchers independently coded each RDP and then compared their results to ensure consistency. 
Typical inconsistencies arose when the same policy used conflicting language in different sections. For example, requiring deposition in a FAIR repository in one place but only \emph{encouraging} it elsewhere. Under our conservative rule, such cases were coded as \emph{encouraged}. Publishers with a single brief, uniform policy across journals (e.g., APS at the time of coding) yielded the highest agreement, whereas more complex, multi-page frameworks with journal-specific appendices (e.g., Frontiers, MDPI, RSC) produced more discrepancies.

\begin{figure*}[t]
    \centering
    \begin{subfigure}[t]{0.48\textwidth}
        \centering
        \includegraphics[width=\linewidth]{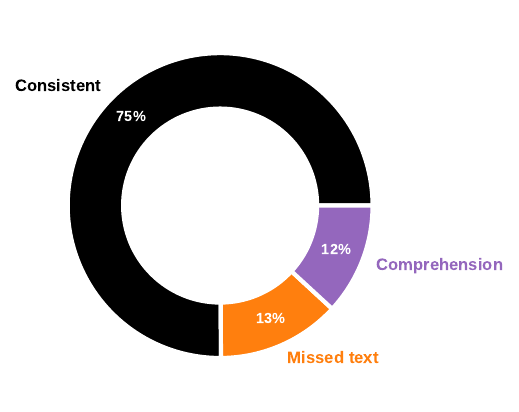}
        \caption{Summary of the percentage of coding questions that were answered consistently by the two encoders per journal and the reason for the inconsistency.}\label{fig:incosistencies}
    \end{subfigure}\hfill
    \begin{subfigure}[t]{0.48\textwidth}
        \centering
        \includegraphics[width=\linewidth]{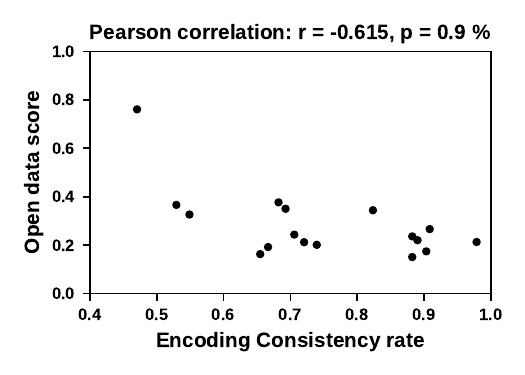}
        \caption{Open data score versus the consistency of encoding, excluding questions 10 \& 11.}\label{fig:ordvsconsistency}
    \end{subfigure}
    \caption{Consistency of encoding and its relationship to open data score.}
    \label{fig:consistency_panels}
\end{figure*}

In Figure~\ref{fig:incosistencies} we provide an overview of the coding process. \SI{75}{\percent} of the questions were initially coded consistently, while initially inconsistent encoded questions could be attributed to RDP text that was missed (\SI{13}{\percent} of all questions) and a misunderstanding of the meaning of the policy (\SI{12}{\percent}).

The most inconsistently-coded items were Questions 10 \& 11 (data types for which unique policies are recommended/required), Question 7 (data sharing method), and Question 6 (timing of data release). Higher inconsistency for Questions 10 \& 11 is expected because these required free-text extraction of data-type–specific rules. For Question 7, the option \emph{``Multiple data sharing methods equally recommended in the RDP''} accounted for \SI{75}{\percent} of its inconsistencies, indicating that this question is ill-posed and should be revised in future studies. For Question 6, distinctions between \emph{``required before publication''} and \emph{``required after publication''} were often unclear in policy text, suggesting that journals could improve the explicitness of release timelines. 

These results led to a \textit{post hoc} hypothesis that policy strictness impacts coding consistency (Figure~\ref{fig:ordvsconsistency}). Using the open data score as a proxy for policy comprehensiveness and excluding questions 10 \& 11 from the consistency rating, we observe a strong negative Pearson correlation of $r=-0.615$ with $p=0.9\,\%$: more comprehensive policies tend to be harder to code consistently. 
As the hypothesis emerged during the analysis phase, these findings should be considered in a hypothesis generating fashion and require confirmation in future work.

\section{Discussion}

Our results show that materials science journals have largely converged on the \emph{language} of open research (e.g., policies and data availability statements), but not on a shared, enforceable baseline that reliably yields reusable datasets and software. The consequence is a fragmented policy landscape in which the practical burden of interpretation and compliance is shifted to authors, editors, and referees. Practically, this means the reproducibility of published work can still hinge on venue-specific norms rather than community-wide expectations.

\noindent
\textbf{Data and Code Policies.} A sizable fraction of journals in materials science (\SI{36.3}{\percent}) mandate data availability by publication, a somewhat higher proportion than reported previously for physics journals~\cite{rousi2020journal}. Nevertheless, a majority of journals still permit delayed or conditional data release, which undermines the goal of immediate reproducibility. Perhaps most concerning are the results regarding \textit{data sharing} itself. Fewer than \SI{7}{\percent} of journals require public deposition of all data, while most journals merely encourage the sharing of data. This disconnect between aspirational policy language and enforceable requirements likely contributes to the persistent gap between journal policy and author practice.

The method of data sharing is an area where improvements can be made at little cost. Encouraging data sharing on FAIR repositories does not require additional resources, and for data types where repositories and standards already exist, journals without a respective policy can easily adopt existing recommendations. Here we observe positive trends. FAIR repositories are frequently encouraged, and citeability is often mentioned, reflecting an awareness of best practices for data management. Still, requirements for FAIR-compliant publication remain rare, and the lack of licensing guidance in the majority of policies (\SI{77.4}{\percent}) raises questions about the reusability of shared data. 

A policy’s practical effect depends on whether it is embedded in editorial routines. While references to data/code in referee guidelines are more common than previously reported~\cite{Resnik2019}, most journals still do not treat data and software as objects that are routinely checked during peer review.
Evaluating whether a dataset actually enables the reproducibility of an article is difficult and unless this is verified during the peer review process, it is unlikely that this can be confirmed later. In our study, we only identified three journals that use a data and/or code referee who is responsible for checking the completeness of the data and confirming the basic functionality of the code.

Coding standards and quality assurance are the cornerstones of industrial software reliability, especially in high-responsibility domains such as the automotive industry~\cite{Parasoft2025} or avionics~\cite{NASA2004}. It is evident that scientific software developers still have a lot to learn from industrial software developers~\cite{Joppa2013}.
Compared to previous studies, Stodden et al.~\cite{stodden2013toward} found a code policy in \SI{22}{\percent} of computational and mathematical biology journals, similar to \SI{19.1}{\percent} for physical science journals in Ref.~\cite{Resnik2019}. Consequently, \SI{70}{\percent} appears as an improvement. However, Resnik et al.~\cite{Resnik2019} also found that \SI{20.1}{\percent} of journals required depositing code into repositories, in stark contrast to \SI{6.4}{\percent} in our dataset. We analyzed the research policies of 17 of the physics science journals that were identified as requiring code deposition in Ref.~\cite{Resnik2019} and found no requirement for code deposition in 11 of the journals. We see two possible explanations for this discrepancy. 
First, RDPs concerning code could have deteriorated. This seems unlikely given the general strengthening of open data requirements reported in our study and others~\cite{rousi2020journal}. A more plausible explanation lies in the challenges of consistently encoding RDPs across and within studies. For example, our study applies a strict interpretation of the term ``required'', as outlined in Section~\ref{sec:methods}, whereas other studies may adopt a more flexible definition. Such differences in interpretation can readily lead to diverging results. In this context, we suggest that future studies should ideally include, as part of their data, the specific text excerpts from RDPs that informed their coding decisions, to better address these types of questions in the future. In addition, we find that even within our own coding process, results varied between encoders, as detailed in Section~\ref{sec:encoding_process} and Section~\ref{SI-sec:reviewing_consistency} in the SI. This variability likely reflects not only deficiencies in the coding framework but also ambiguity in the wording of certain open data policies themselves.

Code quality is almost never mentioned in RDPs, as shown by near-zero \textit{research data scores} of linting and automatic testing. The lack of attention to code quality standards may partly stem from the fact that most materials researchers are not formally trained in software development, and that this is often not the focus of studies. As good software practices require significant time and resources, they are rarely prioritized by funding agencies in materials science. Given the wide diversity of research code --- from short analysis scripts to large-scale simulation packages --- strict and universal standards are difficult to enforce at the journal level. Nevertheless, for journals with a strong focus on method and software development, there is a clear case for requiring higher standards of code quality.

Finally, the inverse correlation between the open data score and encoding consistency (Fig.~\ref{fig:ordvsconsistency}) highlights a tradeoff in policy design: policies that attempt to cover many cases and data types can become harder to interpret and implement. This suggests that strengthening RDPs is not only a question of adding more requirements, but of making requirements \emph{testable} and \emph{unambiguous}. A practical direction is therefore to align policies with existing standards and infrastructure while expressing them as a small set of checks to be verified during the editorial and refereeing process. Concretely, future policy templates could prioritise (i) a concise core with non-conflicting language; (ii) clearly tiered obligations (e.g., \emph{required} $\rightarrow$ \emph{expected} (may affect editorial decisions) $\rightarrow$ \emph{mentioned}), and (iii) a canonical decision path for timing, location (FAIR repositories), persistent identifiers, and licensing, with journal-specific annexes that cannot override the core without explicit precedence rules.

Our results point to a small number of recurrent gaps that are well covered by mature community standards but are rarely made enforceable in journal policies. However, there arelimitations of this study concerning which journals have been chosen for RDP evaluation. Below, we map these gaps to policy clauses and checks that journals could adopt without requiring discipline-specific reinvention.

\subsection*{From observed gaps to policy templates}

\noindent\textbf{(1) From ``encouraged'' to verifiable deposition.}
While many journals encourage repository deposition, only a minority require FAIR-compliant deposition, and licensing guidance is absent in most policies (\SI{77.4}{\percent}). A minimal enforceable baseline is therefore: (i) deposition of all data needed to reproduce the main claims in a community-recognized repository (FAIR where available), (ii) a persistent identifier (typically a DOI), and (iii) an explicit reuse license. These elements are supported by established infrastructure such as DataCite and DOIs~\cite{Brase2009,Neumann2014}. Editorial verification can be lightweight: the data availability statement must contain a resolvable identifier and license, and referees must have confirmed availability of the deposited artifacts at review time.

\noindent\textbf{(2) Treating research software as a first-class research output.}
Our survey shows that code requirements lag far behind data requirements: code sharing is usually optional, and most policies omit dependencies, versions, and reproducible execution environments. A corresponding baseline, aligned with FAIR4RS~\cite{Barker2022Oct_622}, requires an archived versioned code release with a persistent identifier, together with an explicit specification of the execution environment. Here, e.g., \texttt{requirements.txt}/\texttt{environment.yml} are a minimal example for python code that does not enable full reproducibility, while more modern environment managers~\cite{pixi} or a container recipe allow for full reproducibility. Where feasible, journals can require a single ``reproduce'' entry point that regenerates key figures or tables. This is supported by existing repositories and archival services (e.g. Zenodo~\cite{Zenodo}) and by common environment managers (e.g. Conda/Spack/Pixi) and containerization approaches. Importantly, these requirements can be adjusted for scope: journals may apply stricter expectations for methods/software papers than for primarily experimental studies.

\noindent\textbf{(3) Code quality: focusing on minimal signals rather than full industrial compliance.}
Because scientific software varies from short scripts to large simulation packages, universal industry-grade standards are unrealistic at the journal level. However, journals can still require minimal quality signals that improve maintainability and reusability: basic documentation of inputs/outputs, a license, and at least one automated sanity check (unit or functional test) for core routines when software is central to the contribution. Such requirements are consistent with established scientific software engineering guidance~\cite{Wilson2014, Jimenez2017Jun_876} and help address the near absence of current policies.

Given that most journals still do not embed RDP requirements in referee guidance, a key step is procedural: requiring a short reproducibility checklist at submission (data DOI + license, code release + version, environment specification, mapping of figures to artifacts). Where workloads permit, journals can adopt specialized data/code editors for selected submissions, following emerging practice in a small number of venues. This targets the observed gap between policy language and actual verifiability without placing the full burden on individual referees.

\noindent\textbf{(6) Beyond repository-by-repository compliance: interoperability as a future policy dimension.}
Finally, our survey suggests that current RDPs rarely address interoperability across repositories and disciplines. Emerging concepts such as FAIR Digital Objects (FDOs), discussed within the Research Data Alliance and the FDO Forum~\cite{Blumenroehr2024,SoilandReyes2024,Anders2023a,Koers2020}, aim to support machine-actionable resolution of identifiers and access to heterogeneous data across domains. While FDO systems are still developing, the absence of any policy language on cross-domain interoperability indicates an opportunity for forward looking RDP templates to anticipate the increasing reuse of materials data across domains.

\noindent\textbf{(7) Capacity building as part of policy design.}
Finally, stricter and more software-aware policies will only be effective if researchers can realistically comply. Because many materials scientists receive limited formal training in data stewardship and software engineering, journals and publishers can reduce friction by explicitly pointing authors to community training resources (e.g., The Carpentries and CodeRefinery) that instruct on version control, licensing, documentation, testing, continuous integration, and reproducible environments~\cite{Carpentries,CodeRefinery}. Referencing such resources within RDPs complements enforcement with enablement and directly targets the skills gaps that likely contribute to weak code-policy uptake.

\section{Conclusions}
We expanded existing coding frameworks to cover data and code sharing requirements and applied them to create a quantitative survey of materials science research data policies across 171 journals. Overall, the landscape shows clear progress in policy existence and intention, while enforceable requirements remain rare, with research software policies lagging even further behind. Aggregating policy features into an open data score reveals substantial heterogeneity across journals, with journal impact factor and access model providing limited predictive power for policy strength. Finally, double-coding demonstrates that policy complexity and strictness might reduce interpretability, highlighting the need for concise, internally consistent language and clearly tiered obligations.

Together, these findings provide a baseline of current RDP practice in materials science. They motivate evidence-based policy templates that translate community standards into a small set of unambiguous, verifiable checks during submission and peer review, while anticipating future needs such as cross-repository interoperability.

\clearpage
\section{Methods}
\label{sec:methods}
\subsection{Journal Selection}

To select a representative set of journals, we used the \textsc{dimensions}~\cite{dimensions} Python API. As a starting point, we limited our selection to 9 fields of research as defined by the Australian and New Zealand Standard Research Classification (ANZSRC 2020):

\begin{itemize}
    \item 3047 Theoretical and Computational Chemistry
    \item 3403 Macromolecular and Materials Chemistry
    \item 3402 Inorganic Chemistry
    \item 3405 Organic Chemistry
    \item 3406 Physical Chemistry
    \item 4016 Materials Engineering
    \item 4018 Nanotechnology
    \item 5102 Atomic, Molecular, and Optical Physics
    \item 5104 Condensed Matter Physics
\end{itemize} 

Searching for the most popular publishers in terms of the number of articles, we arrived at a preliminary selection. For each publisher, we then selected the three journals with the highest average citation count, the three most popular journals, and three random journals. All available journals were selected for publishers with fewer than nine journals. The detailed query used to obtain the data, along with the resulting dataset, is published with the article. Due to the high time cost of encoding journals, we limited ourselves to the 17 most popular publishers, resulting in a total of 171 journals.

\subsection{Encoding process}
\label{sec:encoding_process}

Two researchers independently coded each publisher’s RDP. For every answer, coders saved the exact policy text that supported their choice (when no relevant policy text existed, this field was necessarily empty). After independent coding, one of the coders compared the two sets to ensure consistency. In cases of discrepancies, the reviewer selected, if possible, the correct choice based on the collected texts, or, if this was not possible, discussed a correct choice with the rest of the research team. The reason for the discrepancy is also noted and analyzed to improve both the coding framework and RDPs for the future. When multiple interpretations were possible or different text sections provided conflicting information, the most conservative interpretation was used. 
The coding process took place between June 2024 and May 2025. For each journal, the date of the encoding is included in the published data. We note that during the study, RDPs already improved, e.g., the research policies of APS became significantly stricter, and we hope to make a systematic comparison of the progress in a few years.

\subsection{Numerical Scoring of Data Policies}\label{sec:ods}
The \textit{open data score} relies on the fact that our encoding framework is exclusive and hierarchically ranked. Each answer $\alpha$ to every question $n$ is assigned a strictness value $w_{n \alpha}$ between 0 (most relaxed) and 1 (most strict). We denote the selected answer of each question with $\alpha_n$. To attain the \textit{open data score} ($s$), we sum over all questions and normalize by the number of questions $N$. 
\begin{equation}
    s = \frac{1}{N} \sum_{n=1}^{N} \sum_{\alpha=1} w_{n \alpha} \delta_{\alpha, \alpha_n}
\end{equation}
Here, $\delta_{\alpha,\alpha_n}$ is the Kronecker delta, indicating whether a given answer was selected.
\subsection{\textbf{Coding framework}}
\label{sec:coding_framework}

To objectively and reproducibly evaluate open data policies, we have developed a coding framework in the form of mutually exclusive multiple-choice questions.  The possible choices are ranked by how strongly they reflect open data principles, with later answers  considered stronger policies. This allows for the association of each question with a numerical value. Our coding framework builds upon previous efforts~\cite{Resnik2019, rousi2020journal}, emphasizing clear rankability and incorporating additional relevant elements, such as scientific code. We mark questions overlapping with Refs.~\cite{Resnik2019} and ~\cite{rousi2020journal} with the respective citations. The coding framework is defined as follows:

\begin{snugshade*}
\begin{enumerate}[left=1mm, label=\textbf{\arabic*.}, noitemsep]
    \item \textbf{Existence of research data policy}~\cite{rousi2020journal}
        \begin{enumerate}[noitemsep]
            \item No
            \item Yes
        \end{enumerate}
    
    \item \textbf{Data sharing requirements} (adapted from~\cite{Resnik2019})
    \begin{enumerate}[noitemsep]
        \item Data sharing only with editors and referees (no public sharing)
        \item Data sharing encouraged but optional
        \item Data sharing required but not public (e.g. available upon request)
        \item Public data sharing required for specific data types
        \item Public data sharing of all data required
    \end{enumerate}
    
    \item \textbf{FAIR data sharing}
    \begin{enumerate}[noitemsep]
        \item Public data sharing on a FAIR repository not mentioned
        \item Public data sharing on a FAIR repository encouraged
        \item Public data sharing on a FAIR repository required for specific data types
        \item Public data sharing of all data on a FAIR repository required
    \end{enumerate}

    \item \textbf{Data availability statement}
    \begin{enumerate}[noitemsep]
        \item Not mentioned
        \item Mentioned but optional
        \item Required
    \end{enumerate}

    \item \textbf{Citability and findability of data}
    \begin{enumerate}[noitemsep]
        \item No mention of DOIs or other persistent identifiers (Handle, Archival Resource Key (ARK), etc.)
        \item DOIs or other persistent identifiers recommended for datasets or code
        \item DOIs or other persistent identifiers required for datasets or code
    \end{enumerate}

    \item \textbf{Timing of data release}\cite{rousi2020journal}
    \begin{enumerate}[noitemsep]
        \item Timing of data availability not addressed
        \item Required data available before review
        \item Required data available before publication
        \item Required data available after publication
        \item Required data available after an embargo period
    \end{enumerate}

    \item \textbf{Recommended data-sharing method} (adapted from~\cite{Resnik2019})
    \begin{enumerate}[noitemsep]
        \item No method recommended
        \item Multiple methods equally recommended
        \item Data sharing upon request to authors
        \item Supplementary material or journal-hosted sharing recommended
        \item Public online repositories recommended
        \item FAIR repositories recommended
    \end{enumerate}

    \item \textbf{Recommended/required licenses}
    \begin{enumerate}[noitemsep]
        \item No mention of licenses
        \item Specific license type mentioned
        \item Open access license required for shared data
        \item Open source license required
    \end{enumerate}

    \item \textbf{Referee guidelines concerning research data}
    \begin{enumerate}[noitemsep]
        \item Data sharing policy not mentioned in refereeing guidelines
        \item Data sharing policy mentioned in refereeing guidelines
        \item Referees confirm shared data/code
        \item Additional data/code-specific referee required
    \end{enumerate}

    \item \textbf{List data types for which unique policies are recommended} (adapted from~\cite{Resnik2019})

    \item \textbf{List data types for which unique policies are required} (adapted from~\cite{Resnik2019})
    
    \item \textbf{Code sharing requirements}
    \begin{enumerate}[noitemsep]
        \item Code sharing not mentioned
        \item Code sharing encouraged but optional
        \item Code sharing required
    \end{enumerate}

    \item \textbf{Code reproducibility}
    \begin{enumerate}[noitemsep]
        \item No mention of dependencies
        \item Listing dependencies and versions encouraged
        \item Listing dependencies and versions required
        \item Container/installation script required
        \item Full reproducibility of figures/tables with container/scripts required
    \end{enumerate}

    \item \textbf{Versioning and persistent identifiers}
    \begin{enumerate}[noitemsep]
        \item No mention of persistent identifiers or versions
        \item Persistent identifier or version encouraged
        \item Persistent identifier or version required
    \end{enumerate}

    \item \textbf{Code quality standards}
    \begin{enumerate}[noitemsep]
        \item No mention of code quality standards
        \item Standards encouraged
        \item Standards required
        \item Specific criteria for referees on code quality
    \end{enumerate}

    \item \textbf{Automatic testing}
    \begin{enumerate}[noitemsep]
        \item No mention of automatic testing
        \item Automatic testing encouraged
        \item Automatic testing required
    \end{enumerate}

    \item \textbf{Code documentation}
    \begin{enumerate}[noitemsep]
        \item No mention of documentation standards
        \item Documentation encouraged
        \item Documentation required
    \end{enumerate}

    \item \textbf{Linting standards}
    \begin{enumerate}[noitemsep]
        \item No mention of linting standards
        \item Linting standards encouraged
        \item Linting standards required
    \end{enumerate}

    \item\textbf{Code development and continuous integration}
    \begin{enumerate}[noitemsep]
        \item No mention of development standards
        \item Development standards encouraged
        \item Development standards required
    \end{enumerate}
\end{enumerate}
\end{snugshade*}

\section{Acknowledgments}
This article/publication is a result of joint work in COST Action CA22154 - Data-driven Applications towards the Engineering of functional Materials: an Open Network (DAEMON) supported by COST (European Cooperation in Science and Technology).
JS was supported by the European Research Council (ERC) under the European Union’s Horizon 2020 research and innovation program project HERO Grant Agreement No. 810451 and funded by the SNSF Ambizione grant number 233444.
LH was supported by a UKRI Horizon grant (MSCA, EP/Y024923/1). 
KB acknowledges the support by Ministry of Science, Technology and Innovation of the Republic of Serbia via contract no. 451-03-33/2026-03/ 200017.
MS acknowledges support by the European Union by ERC grant, project no. 101161287.
The views and opinions expressed are, however, those of the authors only and do not necessarily reflect those of the European Union or the European Research Council Executive Agency. Neither the European Union nor the granting authority can be held responsible for them.
SB acknowledges support from the National Science Center in Poland through the SONATA BIS grant DEC-2023/50/E/ST3/00569 and from the Foundation for Polish Science in Poland through the FIRST TEAM FENG.02.02-IP.05-0177/23 project.
All authors gratefully acknowledge Nicola Spaldin, Julian Dederke and Nicolai Bissantz for helpful and insightful discussions, and the ETH Library for their support.

\section{Author Contributions}
Conceptualization: FG, HM, JS, LH, MS, RKSAH, SG; Methodology: HM, JS, LH, MS, SG; Software: HM, IL, JS, LH, MS, NL, SG; Validation: FG, JS, LH, MS, RKSAH, SG; Formal analysis: FG, HM, JS, LH, MS, SG; Investigation: BK, ES, FG, HM, IL, JMO, JS, KB, KK, KR, LH, MS, MT, PP, PR, RKSAH, SB, SG, SPS; Data Curation: FG, HM, JS, LH, NL, SG; Writing – Original Draft: FG, HM, JS, LH, MS, SG; Writing – Review \& Editing: BK, ES, FG, HM, IL, JMO, JS, KB, KK, KR, LH, MS, MT, PP, PR, RKSAH, SB, SG, SPS; Visualization: FG, HM, JS, LH, MS; Supervision: FG, JS, LH, MS; Project Administration: JS, LH, MS; Funding Acquisition: FG, KR, MT

\section{Competing  Interests}
KR is part of the editorial board of Journal of Physics:Materials (IOP Publishing)

\section{Data and Code Availability}
All data and code to reproduce the research are available at \url{https://github.com/daemoncost/Journal-Research-Data-Policy}. The code version at publication time will be mirrored to Zenodo with a persistent identifier.

\bibliography{main}

\end{document}